\documentstyle[amssymb,prl,aps,epsfig]{revtex}

\begin{document}

\title{Probing thermal fluctuations and inhomogeneities in type II superconductors by means
of applied magnetic fields}
\author{T. Schneider \\
Physik-Institut der Universit\"{a}t Z\"{u}rich, Winterthurerstrasse 190,\\
CH-8057 Z\"{u}rich, Switzerland}
\date{}
\maketitle

\begin{abstract}
A superconductor is influenced by an applied magnetic field. Close
to the transition temperature $T_{c}$ fluctuations dominate and
the correlation length $\xi $ increases strongly when $T_{c}$ is
approached. However, for nonzero magnetic field $H$ there is
another length scale $L_{H}=\sqrt{\Phi _{0}/aH}$ where $a$ is a
universal amplitude. It is comparable to the average distance
between vortex lines. We show that the correlation length is
bounded by this length scale, so that $\xi $ cannot grow beyond
$L_{H}$. This implies that type II superconductors in a magnetic
field do not undergo a phase transition to a state with zero
resistance. We sketch the scaling theory of the resulting magnetic
field induced finite size effect. In contrast to its inhomogeneity
induced counterpart, the magnetic length scale can be varied
continuously in terms of the magnetic field strength. This opens
the possibility to assess the importance of fluctuations, to
extract critical point properties of the homogeneous system and to
derive a lower bound for the length scale of inhomogeneities which
affect thermodynamic properties. Our analysis of specific heat
data for under- and optimally doped YBa$_{2}$Cu$_{3}$O$_{7-\delta
}$, MgB$_{2}$, 2H-NbSe$_{2}$ and Nb$_{77} $Zr$_{23}$ confirms this
expectation. The resulting lower bounds are in the range from 182
A to 818 A. Since the available data does not extend to low
fields, much larger values are conceivable. This raises serious
doubts on the relevance of the nanoscale spatial variations in the
electronic characteristics observed with scanning tunnelling
microscopy.

\end{abstract}

\bigskip

\bigskip
\section{Introduction}

A primary effect of a magnetic field on a type II superconductor
is to shift the specific heat peak to a lower temperature. Indeed,
this shift provided some of the earliest evidence for type II
superconductivity\cite{morin}. In a mean-field treatment this
behavior can be comprehended as follows: To retain the continuous
character of the zero field transition the Ginzburg-Landau
correlation length $\xi =\xi _{0}\left| 1-T/T_{c}\right| ^{-1/2}$
is supposed to diverge along the so called upper critical line of
continuous phase transitions defined by
\begin{equation}
t_{c2}\left( H\right) =1-\frac{T_{c2}\left( H\right)
}{T_{c}}=\frac{2\pi \xi _{0}^{2}H}{\Phi _{0}},\ \ H_{c2}\left(
T\right) =\frac{\Phi _{0}}{2\pi \xi _{0}^{2}}\left(
1-\frac{T}{T_{c}}\right) .  \label{eq1}
\end{equation}
Calculations of the specific heat within the framework of the
Abrikosov theory and its generalizations\cite{fetter} yield at
this mixed state to normal state phase boundary for the jump in
the specific heat coefficient the expression $\Delta c\left(
H_{c2},T_{c2}\right) /T_{c2}\propto \left. \left(
dH_{c2}/dT\right) ^{2}\right| _{T_{c2}}$, where the factor of
proportionality is of order unity and takes the effect of the
vortex lattice into account. Matching with the zero field jump
then requires that
\begin{equation}
\frac{\Delta c\left( H_{c2},T_{c2}\right) /T_{c2}}{\Delta c\left(
T_{c}\right) /T_{c}}\propto \frac{T_{c2}\left( H\right)
}{T_{c}}=1-\frac{2\pi \xi _{0}^{2}H}{\Phi _{0}}.  \label{eq2}
\end{equation}
Thus, the magnetic field reduces both, the transition temperature
and the jump in the specific heat. Although this mean-field
scenario describes the experimental data of various type II
superconductors with comparatively large correlation length
amplitude $\xi _{0}$ , including \ MgB$_{2}$\cite {lyard},
2H-NbSe$_{2}$\cite{sanchez}, and Nb$_{77}$Zr$_{23}$\cite
{mirmelstein}, reasonably well, it neglects an essential effect of
an applied magnetic field. In analogy to liquid $^{4}$He in a
uniformly rotating container\cite{hauss}, the magnetic field
applied to a type II superconductor is an external perturbation
which drives the system away from criticality. Close to the
transition temperature $T_{c}$ the correlation length $\xi $
increases strongly when $T_{c}$ is approached. However, for
nonzero rotation frequency $\Omega $ or magnetic field $H$ the
average distance between vortex lines , $L_{\Omega }\approx
\sqrt{\left( \pi \hbar \right) /m_{4}\Omega }\ $\ or $L_{H}\approx
\sqrt{\Phi _{0}/\left( H\right) } $ is a limiting length scale.
For this reason $\xi $ cannot grow beyond $L_{\Omega ,H}$.
Consequently, for finite $\Omega $ or $H$, the thermodynamic
quantities like the specific heat are smooth functions of
temperature. As a remnant of the singularities along $T_{c2}\left(
H\right) $ the quantities exhibit a maximum or an inflection point
along $T_{p}\left( H\right) $. Such a finite size effect also
occurs in zero magnetic field due to inhomogeneities with length
scale $L$.

In this context it is instructive to refer to the early literature
focused on fluctuation effects in type II superconductors.
Implicit in most of these treatments is the assumption that
fluctuations do not interact; that is, only Gaussian fluctuations
are considered\cite{lee}. Although this assumption breaks down in
the critical region, it captures some of the qualitative aspects
of fluctuations in zero magnetic field. If the Gaussian
approximation is used to calculate thermodynamic properties of a
type II superconductor near the mean-field phase boundary
$T_{c2}\left( H\right) $, singular behavior of thermodynamic and
transport properties is predicted to occur\cite{lee}. However, the
fluctuations of a bulk superconductor in a magnetic field become
effectively one dimensional, as noted by Lee and Shenoy
\cite{lee}; and a bulk superconductor behaves like an array of
rods parallel to the magnetic field with diameter $L_{H}$. Since
fluctuations become more important with reduced dimensionality and
there is a limiting magnetic length scale $L_{H}$ it becomes clear
that the critical line $T_{c2}\left( H\right) $ is an artefact of
the approximations. Indeed, calculations of the specific heat in a
magnetic field which treat the interaction terms in the Hartree
approximation and extensions thereof, find that the specific heat
is smooth through the mean-field transition temperature
$T_{c2}\left( H\right) $\cite{thouless,brezin,hikami}. In the
context of finite size scaling this is simply due to the fact that
the correlation length of fluctuations which are transverse to the
applied magnetic field are bounded by the magnetic length
$L_{H}\propto H^{-1/2}$.

A main point of this paper is to show that in a type II
superconductor the correlation length cannot grow beyond the
magnetic length scale
\begin{equation}
L_{H}=\sqrt{\frac{\Phi _{0}}{aH}},  \label{eq5}
\end{equation}
where $a$ is a universal amplitude when critical fluctuations
dominate. Hence, the correlation length $\xi ^{-}\left( t\right) $
is bounded by
\begin{equation}
\xi ^{-}\left( t\right) =\xi _{0}^{-}\left| t\right| ^{-\nu
}=L_{H}, \label{eq6}
\end{equation}
where $\nu $ is the critical exponent of the correlation length.
Beyond the mean-field approximation it differs from $\nu =1/2$.
Due to this finite size effect the maximum of the specific heat
peak is shifted to a lower temperature $T_{p}$ given by
\begin{equation}
t_{p}\left( H\right) =1-\frac{T_{p}\left( H\right) }{T_{c}}=\left(
\frac{\xi _{0}^{-}}{L_{H}}\right) ^{1/\nu }=\left( \frac{a\left(
\xi _{0}^{-}\right) ^{2}H}{\Phi _{0}}\right) ^{1/2\nu }.
\label{eq7}
\end{equation}
Although this finite size scaling result agrees formally with the
mean-field expression for $t_{c2}$ (Eq.(\ref{eq1})) with $\nu
=1/2$ and $a=2\pi $, $t_{p}\left( H\right) $ is not a line of
continuous phase transitions. Along this line the specific heat
peak adopts its maximum value and the correlation length attains
the limiting length scale $L_{H}$ . On the other hand, there are
sample inhomogeneities with length scale $L$, as well. They imply
that the correlation length $\xi ^{-}$ cannot grow beyond $L$,
\begin{equation}
\xi ^{-}\left( t\right) =L.  \label{eq9}
\end{equation}
Thus, the zero-field specific heat singularity is rounded.
Replacing in Eq.(\ref{eq7}) $L_{H}$ by $L$, it is seen that the
maximum of the zero field specific heat shifts to the lower
temperature $T_{p}\left( L\right) $. However, in contrast to the
magnetic counterpart, where the length scale $L_{H}\propto
H^{-1/2}$ can be varied at will in terms of the magnetic field
strength, the length scale of the inhomogeneities remains fixed.
In type II superconductors where inhomogeneities are unavoidable,
two limiting regimes, characterized by
\begin{equation}
L_{H}<L,\ \ L_{H}>L,  \label{eq10}
\end{equation}
can be distinguished. For $L_{H}<L$ the magnetic field induced
finite size effect limits $\xi $ to grow beyond $L_{H}$, while for
$L_{H}>L$ it is the length scale of the inhomogeneities. Since
$L_{H}$ can be tuned by the strength of the applied magnetic field
(Eq.(\ref{eq5})), both limits are experimentally accessible.
$L_{H}<L$ is satisfied for sufficiently high and $L_{H}>L$ for low
magnetic fields. Thus, the occurrence of a magnetic field induced
finite size effect requires that the magnetic field and the length
scale of inhomogeneities affecting thermodynamic properties
satisfy the lower bounds
\begin{equation}
H>\frac{\Phi _{0}}{aL^{2}},\ \ L>\sqrt{\frac{\Phi _{0}}{aH}}.
\label{eq11}
\end{equation}
It is important to recognize that these bounds hold, whenever
fluctuations are at work. Noting that thermodynamic and transport
measurements are usually performed up to $12$ T, it becomes clear
that the magnetic field induced finite size effect allows to trace
fluctuations, to extract critical properties of the homogeneous
system and to derive a lower bound for the length scale of
inhomogeneities which affect the thermodynamic properties of type
II superconductors. Such an analysis may also help to distinguish
between possible mechanisms where the length scale of the
inhomogeneities enters in an essential way. Below, we give a
summary of the main results of this paper. A related analysis of
expansivity measurements on YBa$_{2}$Cu$_{3}$O$_{7-\delta }$
exposed to a magnetic field was recently performed by
Lortz\cite{lortz}, in terms of a comparison with a model for
uniformly rotating $^{4}$He\cite{hauss}.

We show that in isotropic extreme type II superconductors and
nonzero magnetic field $H$ the correlation length cannot grow
beyond the limiting length scale $L_{H}=\sqrt{\Phi _{0}/\left(
aH\right) }$, where $a$ is a universal amplitude. Since $a$ turns
out to be of order one, $L_{H}$ is comparable to the average
distance between vortex lines. Given this limiting length scale,
we derive the scaling expression for the line $T_{p}\left(
H\right) $, along which the correlation length attains the
limiting magnetic length scale $L_{H}$. It turns out that close to
zero field criticality the ratio between $T_{p}\left( H\right) $
and the melting line $T_{m}\left( H\right) $ is a universal
number. These results, extended to anisotropic type II
superconductors are found to be in good agreement with the
experimental results of Schilling \emph{et al}.\cite{schilling}
and Roulin \emph{et al}.\cite{roulin} for
YBa$_{2}$Cu$_{3}$O$_{7-\delta }$. Moreover, we find that the data
of Roulin \emph{et al}.\cite{roulin} for the temperature and
magnetic field dependent specific heat coefficient, properly
rescaled, falls on a single universal curve, which is the finite
size scaling function. The phase diagram which emerges consists of
the lines $T_{p}\left( H_{i}\right) $ and $T_{m}\left(
H_{i}\right) $, where $H_{i}$ is the $i$-th component of the
applied magnetic field $\mathbf{H}$. Our analysis of the data for
the optimally doped sample\cite{roulin} provides consistent
evidence for critical properties falling into 3D-XY universality
class\cite{hohenberg,privman,peliasetto}, while the data of the
underdoped sample\cite{junod} uncovers the characteristics of a
3D-2D crossover\cite {book,parks}. Although the mean-field
scenario describes the experimental data of various type II
superconductors with comparatively large correlation length
amplitudes, including \ MgB$_{2}$\cite{lyard}, 2H-NbSe$_{2}$\cite
{sanchez} and Nb$_{77}$Zr$_{23}$\cite{mirmelstein}, remarkably
well, we find that this treatment underestimates the relevance of
fluctuations. Indeed the experimental data for the specific heat
coefficient is fully consistent with a magnetic field induced
finite size effect, although the critical regime is not attained.
As a by-product we deduce for the length scale of the
inhomogeneities affecting the thermodynamic properties of \
YBa$_{2}$Cu$_{3}$O$_{7-\delta }$, MgB$_{2}$, 2H-NbSe$_{2}$ and
Nb$_{77}$Zr$_{23}$ lower bounds ranging from $182$ A to $814$ A.
Surprisingly enough, $814$ A applies to the cubic superconducting
alloy Nb$_{77}$Zr$_{23}$. Since the available data does not extend
to very low fields larger values are conceivable. In any case,
these lower bounds are consistent with the estimates obtained from
a finite size scaling analysis of \ the zero field specific heat
data of nearly optimally doped YBa$_{2}$Cu$_{3}$O$_{7-\delta }$
single crystals, giving length scales ranging from 290 to 419 A
\cite{book,housten}. This raises serious doubts on the relevance
of the nanoscale spatial variations in the electronic
characteristics observed in underdoped Bi-2212 with scanning
tunnelling microscopy\cite{liu,chang,cren,lang}. As STM is a
surface probe, it appears to be that these nanoscale
inhomogeneities represent a surface property. Furthermore, in
spite of the consensus that in type II superconductors in an
applied magnetic field a phase transition to a state of zero
resistance should occur, we show that the magnetic finite size
effect points to the opposite conclusion. This is in agreement
with the work reported in Refs.\cite{strachan} and \cite{lobb}.

The paper is organized as follows: In Sec.II we present a short
sketch of the finite size scaling theory\cite{fisher,privf} and
show that in isotropic extreme type II superconductors and nonzero
magnetic field $H$ the correlation length cannot grow beyond the
limiting length scale $L_{H}=\sqrt{\Phi _{0}/\left( aH\right) }$,
where $a$ is a universal constant. Since $a$ turns out to be of
order one, $L_{H}$ is comparable to the average distance between
vortex lines. Given this magnetic field induced limiting length
scale, we derive the scaling expression for the line $T_{p}\left(
H\right) $, along which the correlation length attains the
limiting magnetic length scale $L_{H}$. It turns out that close to
zero field criticality the ratio between $T_{p}\left( H\right) $
and the melting line $T_{m}\left( H\right) $ is a universal
number. These results are then extended to anisotropic type II
superconductors. Moreover, we find that the data for temperature
and magnetic field dependent specific heat coefficient, properly
rescaled, should fall on a single universal curve, which
corresponds to the finite size scaling function. In Sec.III we
analyze experimental data for the temperature dependence of the
specific heat coefficient in various magnetic fields in terms of
the inhomogeneity and magnetic field induced finite size effects.
We concentrate on under- and optimally doped
YBa$_{2}$Cu$_{3}$O$_{7-\delta }$, where single crystal data for
fields applied parallel and perpendicular to the ab-plane are
available \cite{schilling,roulin,junod} and extend to the critical
regime. Accordingly, the full potential of the magnetic field
induced finite size effect can be explored. The analysis of the
data for the optimally doped sample provides consistent evidence
for critical properties falling into 3D-XY universality class,
while the data of the underdoped sample uncovers the
characteristics of a 3D-2D crossover\cite {book,parks}. Although
the mean-field scenario describes the experimental data of various
type II superconductors with comparatively large correlation
length amplitudes, including \ MgB$_{2}$\cite{lyard},
2H-NbSe$_{2}$\cite {sanchez} and
Nb$_{77}$Zr$_{23}$\cite{mirmelstein}, remarkably well, we find
that this treatment underestimates the relevance of fluctuations.
Indeed, the experimental data for the specific heat coefficient is
fully consistent with a magnetic field induced finite size effect,
although the critical regime is not attained. As a by-product we
obtain for the length scale of the inhomogeneities affecting the
thermodynamic properties of \ YBa$_{2}$Cu$_{3}$O$_{7-\delta }$,
MgB$_{2}$, 2H-NbSe$_{2}$ and Nb$_{77}$Zr$_{23}$ lower bounds
ranging from $182$ A to $814$ A. Surprisingly enough, $814$ A
applies to the cubic superconducting alloy Nb$_{77}$Zr$_{23}$.
Since the available data does not extend to very low fields larger
values are conceivable. This raises serious doubts on the
relevance of the nanoscale spatial variations in the electronic
characteristics observed in underdoped Bi-2212 with scanning
tunnelling microscopy\cite{liu,chang,cren,lang}. Although we
concentrate on the specific heat, it is shown that the magnetic
field induced finite size effect also affects the other
thermodynamic and the transport properties, including the
magnetoconductivity.
\section{Sketch of the scaling theory for the magnetic field induced finite size effect}

The finite size scaling theory is based on the assumption that the
system feels its finite size when the correlation length becomes
of the order of the confining length $L$. For a physical quantity
$Q$ this statement can be expressed as\cite{fisher,privf}
\begin{equation}
Q\left( t,L\right) =Q\left( t,L=\infty \right) \ f\left( x\right)
, \label{eq12}
\end{equation}
where
\begin{equation}
x=\frac{L}{\xi \left( t,L=\infty \right) },\ \
t=\frac{T}{T_{c}}-1. \label{eq13}
\end{equation}
$L$ denotes the relevant confining length, $T_{c}$ the transition
temperature and $\xi \left( t,L=\infty \right) $ the correlation
length of the bulk system. Thus, the correlation length, $\xi
^{\pm }\left( t,L=\infty \right) =\xi ^{\pm }=\xi _{0}^{\pm
}\left| t\right| ^{-\nu }$, with critical amplitude $\xi _{0}^{\pm
}$, critical exponent $\nu $, where $\pm =\mathrm{sign}(t)$,
cannot grow beyond $L$ as $t\rightarrow 0$. Thus at $T=T_{p}$,
where
\begin{equation}
\xi ^{\pm }\left( T_{p}\right) =L,  \label{eq14}
\end{equation}
there is no singularity and the transition is rounded. Due to this
finite size effect the maximum of the specific heat peak occurs at
a temperature $T_{p}$ shifted by
\begin{equation}
\left| t_{p}\right| =\left| \frac{T_{p}}{T_{c}}-1\right| =\left(
\frac{\xi _{0}^{-}}{L}\right) ^{1/\nu }  \label{eq15}
\end{equation}
Given the singular behavior of the specific heat coefficient of
the perfect system
\begin{equation}
\frac{c}{T}=\frac{A^{\pm }}{\alpha }\left| t\right| ^{-\alpha
}+B^{\pm }=\frac{1}{\alpha }\left( \frac{R^{\pm }}{\xi _{0}^{\pm
}}\right) ^{3}\left| t\right| ^{-\alpha }+B^{\pm },  \label{eq16}
\end{equation}
the magnitude of the peak located at $T_{p}$ scales then as
\begin{equation}
\frac{c\left( t_{p}\right) }{T_{p}}=\frac{A^{-}}{\alpha }\left|
t_{p}\right| ^{-\alpha /\nu }+B^{-}=\frac{1}{\alpha }\frac{\left(
R^{-}\right) ^{3}}{V_{c}^{-}}\left( \frac{\xi _{0}^{-}}{L}\right)
^{-\alpha /\nu }+B^{\pm }. \label{eq17}
\end{equation}
Here we used the universal relation\cite{hohenberg,privman}
\begin{equation}
A^{\pm }V_{c}=\left( R^{\pm }\right) ^{3},\ V_{c}^{\pm }=\left(
\xi _{0}^{\pm }\right) ^{3}  \label{eq18}
\end{equation}
between the critical amplitudes of the specific heat coefficient
$A^{\pm }$, correlation length and correlation volume $V_{c}$.
$\alpha $ is the critical exponent of the specific heat
singularity and $R^{\pm }$ are universal numbers. Since
superconductors fall in the experimentally accessible critical
regime into the 3D-XY universality class, with known critical
exponents and critical amplitude combinations, we take the
universal properties, including\cite{peliasetto}
\begin{equation}
\alpha =2-3\nu =-0.013,\ \nu =0.671,\ \frac{A^{+}}{A^{-}}=1.07,\
R^{-}=0.815, \label{eq19}
\end{equation}
for granted. In superconductors the order parameter is a complex
scalar for which one has to define below $T_{c}$ the transverse
correlation length $\xi ^{T}$. It is proportional to the helicity
modulus, related to the London penetration depth
via\cite{hohenberg,privman,book}
\begin{equation}
\xi ^{T}=\frac{16\pi ^{3}k_{B}T\lambda ^{2}}{\Phi _{0}^{2}}.
\label{eq19a}
\end{equation}

Since real systems are not only inhomogeneous but are almost
always impure, it is important to clarify whether or not quenched
disorder affects the critical behavior. In general, the critical
behavior of disordered systems is complicated, but there is a
criterion, due to Harris \cite{harris}. It states that the
critical behavior of disordered systems does not differ from that
of the pure system when $\alpha <0$. This is the case when the
critical exponent $\alpha $ of the specific heat is negative.

Rewriting Eq.(\ref{eq17}) in the form
\begin{equation}
\left( \frac{c\left( T_{p}\right) }{T_{p}}-B^{-}\right) \left(
\frac{\xi _{0}^{T}}{L}\right) ^{\alpha /\nu }=\frac{A^{-}}{\alpha
}=\frac{1}{\alpha }\frac{\left( R^{-}\right) ^{3}}{V_{c}^{-}},\
V_{c}^{-}=\left( \xi _{0}^{T}\right) ^{3}  \label{eq20}
\end{equation}
we realize that the $L$ dependence of the specific heat
coefficient at $T_{p} $ allows to determine critical amplitude and
exponent combinations. A useful generalization of this relation is
obtained from Eqs.(\ref{eq12}) and (\ref{eq13}),
namely\cite{schultka,schultka2}
\begin{equation}
\left( \frac{c\left( T\right) }{T}-B^{-}\right) /\left(
\frac{A^{-}}{\alpha }\left| t\right| ^{-\alpha }\right) =g\left(
y\right) ,\ y=x^{1/\nu }=\left( \frac{L}{\xi }\right) ^{1/\nu
}=t\left( \frac{L}{\xi _{0}}\right) ^{1/\nu }, \label{eq21}
\end{equation}
where $g\left( y\right) $ is the finite size scaling function of
its argument. In the limit $L\rightarrow \infty $ and at small but
fixed values of $t$, $g\left( \pm y=\infty \right) =1$. When we
leave $L$ fixed and consider the limit $\pm t\rightarrow 0$, we
obtain $g\left( y\right) \propto \left| y\right| ^{\alpha }$,
while at $T_{p}$, where $y=y_{p}=t_{p}\left( L/\xi _{0}^{T}\right)
^{1/\nu }=-1$, Eq.(\ref{eq17}) requires that $g\left( -1\right)
=1$. This scaling form allows to determine how far the specific
heat of a system with restricted extent can be described in terms
of a universal scaling function.

The study of finite size effects in the $\lambda $-transition of
$^{4}$He, as the bulk is confined more and more tightly in one or
more dimensions\cite {schultka,schultka2}, as well as computer
simulations of model systems \cite {privf}, have been a very good
ground for testing such finite size scaling predictions. In this
context it is essential to recognize that a finite size scaling
analysis allows to assess the importance of critical fluctuations
and to extract universal critical point properties of the
homogeneous system. Moreover, a systematic analysis also allows to
determine the length scale and even the shape of the
inhomogeneities\cite{schultka,schultka2}.

In type II superconductors and nonzero magnetic field $H$ there is
an additional limiting length scale. To derive this magnetic
length scale we note that the free energy density of an
anisotropic type II superconductor adopts below but close to the
zero field transition temperature\ $T_{c}$ the scaling
form\cite{book,tshk,schneieuro,hoferprb}
\begin{equation}
f_{s}={\frac{k_{B}TQ_{3}^{-}}{\xi _{x}^{T}\xi _{y}^{T}\xi
_{z}^{T}}}G_{3}^{-}(z),\quad G_{3}^{-}(0)=1,\ z={\frac{1}{\Phi
_{0}}}\left( \left( \xi _{x}^{T}H_{x}\right) ^{2}+\left( \xi
_{y}^{T}H_{y}\right) ^{2}+\left( \xi _{z}^{T}H_{z}\right)
^{2}\right) ^{1/2}.  \label{eq22}
\end{equation}
$\xi _{i}^{T}$ is the transverse correlation length along
direction $i$, diverging in zero field as $\xi _{i}^{T}=\xi
_{i,0}^{T}|t|^{-\nu }$, $Q_{3}^{-}$ is a universal number
\cite{hohenberg,privman} and $G_{3}^{-}(z)$ a universal scaling
function of its argument. Furthermore, $\pm =\mathrm{sign}(t)$,
$t=T/T_{c}-1$ and $\mathbf{H}=(H_{x},H_{y},H_{z})$ is the applied
magnetic field. In zero field (i.e. $z=0$) one recovers for
$G^{-}(0)=1$ the scaling expression of the 3D XY universality
class\cite{hohenberg,privman}, leading to the universal relation
(\ref{eq18}). In the isotropic case ($\xi _{x}^{T}=\xi
_{y}^{T}=\xi _{z}^{T}$) this scaling form was confirmed by
perturbation theory\cite{lawrie}. This scaling should give an
adequate description of the physics of extreme type II single
crystal superconductors in an intermediate critical regime, where
gauge fluctuations are suppressed, leaving a quenched vector
potential. In establishing the limiting magnetic length scale we
consider the isotropic case, to simplify the notation. In analogy
to uniformly rotating superfluid $^{4}$He it does not depend on
temperature\cite{hauss}. For this reason we set $T=T_{c}$ so that
in zero field $\xi ^{T}\left( T_{c},H=0\right) =\infty $ and for
small magnetic fields the scaling variable , while for low
magnetic fields the scaling variable $z=z=\xi ^{T}H^{2}/\Phi _{0}$
tends to infinity. In this limit $G^{-}\left( z\right) $ adopts
the form\cite{book,schneieuro}
\begin{equation}
G_{3}^{-}(z\rightarrow \infty )=\frac{2C_{\infty }^{-}}{3}z^{3/2}.
\label{eq23}
\end{equation}
where $C_{\infty }^{-}$ is a universal constant. Substitution into
Eq.(\ref {eq22}) yields for the singular part of the free energy
density the expression
\begin{equation}
f_{s}={\frac{Q_{3}^{-}k_{B}T_{c}}{\left( \xi ^{T}\right)
^{3}}}G_{3}^{-}(z)\rightarrow {\frac{2}{3}}Q_{3}^{-}C_{\infty
}^{-}k_{B}T_{c}\left( \frac{H}{\Phi _{0}}\right)
^{3/2}=Q_{3}^{-}\frac{k_{B}T_{c}}{L_{H}^{3}}=Q_{3}^{-}\frac{k_{B}T_{c}}{\left(
\xi ^{T}\left( T_{c},H\right) \right) ^{2}},  \label{eq24}
\end{equation}
where
\begin{equation}
L_{H}=\xi ^{T}\left( T_{c},H\right) =\left( \frac{\Phi
_{0}}{aH}\right) ^{1/2},\ \ a=\left( \frac{2C_{\infty
}^{-}}{3}\right) ^{2/3},  \label{eq24a}
\end{equation}
and $a$ is a universal constant. This relationship between the
limiting magnetic length scale and the transverse correlation
length is also obtained from the scaling form
\begin{equation}
\left( \xi ^{\pm }\left( T,H\right) \right) ^{-2}=\frac{\left|
t\right| ^{2\nu }}{\left( \xi _{0}^{T}\right) ^{2}}S\left(
z\right) =\frac{H}{z\Phi _{0}}S^{-}\left( z\right) ,\ \
S^{-}\left( z=0\right) =1.  \label{eq25}
\end{equation}
The divergence of $\xi ^{T}\left( T_{c},H\right) $ for
$H\rightarrow 0$ requires,
\begin{equation}
S^{-}\left( z\rightarrow \infty \right) =\widetilde{a}z,
\label{eq26}
\end{equation}
so that
\begin{equation}
\left( \xi ^{T}\left( T_{c},H\right) \right)
^{2}=\frac{\widetilde{a}H}{\Phi
_{0}}=\frac{\widetilde{a}}{a}L_{H}^{2}=L_{H}^{2},  \label{eq27}
\end{equation}
because  $\widetilde{a}/a=1$ follows from Eq.(\ref{eq24}),
rewritten in the form $f_{s}/\left( k_{B}T_{c}Q_{3}^{-}\right)
=\xi ^{T}\left( T_{c},H\right) ^{-3\text{ }}=L_{H}^{-3}$.{\ Noting
that the limiting magnetic length scale } $L_{H}$ does not depend
on temperature it applies above and below the zero field
transition temperature. In particular, below $T_{c}$ the
transverse correlation length $\xi ^{T}\left( T,H\right) $ is
bounded by $L_{H}$,
\begin{equation}
\xi ^{T}\left( T,H\right) \leq L_{H}.  \label{eq28}
\end{equation}
This implies that in the $\left( H,T\right) $-phase diagram there
is no line of continuous phase transitions requiring $\xi
^{T}\left( T,H\right) $ to diverge. Furthermore, in the presence
of a weak magnetic field the thermodynamic property, $O$, of a
type II superconductor takes below $T_{c\text{ }}$the scaling form
\begin{equation}
Q\left( t,H\right) =Q\left( t,H=0\right) \ G_{Q}^{-}\left(
z\right) ,..\ z=\frac{aH\left( \xi ^{T}\right) ^{2}}{\Phi
_{0}}=\left( \frac{\xi ^{T}}{L_{H}}\right) ^{2}.  \label{eq28a}
\end{equation}
It establishes the connection to the scaling form (\ref{eq12}) for
a system in a confined geometry of length $L$.

In practice there are also inhomogeneities of length scale $L$. As
aforementioned, two limiting regimes characterized by $L_{H}<L,\ \
L_{H}>L$ (Eq.(\ref{eq10})) can be distinguished. For $L_{H}<L$ the
magnetic field induced finite size effect sets the limiting length
scale, because $\xi ^{T}\left( T,H\right) $ is bounded by $L_{H}$,
while for $L_{H}>L$, $\xi ^{T}\left( T,H\right) $ is limited by
$L$, the length scale of the inhomogeneities. Since
$L_{H}=\sqrt{\Phi _{0}/\left( aH\right) }$ (Eq.(\ref {eq24a})),
both limits are experimentally accessible. $L_{H}<L$ requires
sufficiently large and $L_{H}>L$ small magnetic fields. Moreover,
the occurrence of a magnetic field induced finite size effect
requires that the magnetic field strength exceeds the lower bound
$H>\Phi _{0}/\left( aL^{2}\right) $(Eq.(\ref{eq11})). On the
contrary its absence up to the field $H$ provides a lower bound
for the length scale of the inhomogeneities $L>\sqrt{\Phi
_{0}/\left( aH\right) }$ (Eq.(\ref{eq24a})).

Supposing that $L_{H}<L$ is satisfied, the singularity in the
specific heat is replaced by a broad peak reaching its maximum
value at $T_{p}<T_{c}$ which decreases with increasing field
strength. From Eq.(\ref{eq15}) we obtain for the shift and with
Eq.(\ref{eq17}) for the height of the peak in the specific heat
coefficient the expressions
\begin{equation}
\left| t_{p}\right| =\left| \frac{T_{p}}{T_{c}}-1\right| =\left(
\frac{\xi _{0}^{T}}{L_{H}}\right) ^{1/\nu }=\left( \frac{\left(
\xi _{0}^{T}\right) ^{2}a}{\Phi _{0}}\right) ^{1/2\nu }H^{1/2\nu
},  \label{eq29}
\end{equation}
and
\begin{equation}
\frac{c\left( T_{p}\right) }{T_{p}}=\frac{A^{-}}{\alpha }\left|
t_{p}\right| ^{-\alpha }+B^{-}=\frac{\left( R^{-}\right)
^{3}}{\alpha }\left( \xi _{0}^{T}\right) ^{-3-\alpha /\nu }\left(
\frac{aH}{\Phi _{0}}\right) ^{-\alpha /2\nu }+B^{-},  \label{eq30}
\end{equation}
respectively. We note that for $\alpha <0$ (Eq.(\ref{eq19})) the
peak height decreases monotonically even for small fields. Since
for fixed magnetic field strength $H$, $\left| t_{p}\right|
\propto \left( \xi _{0}^{T}\right) ^{1/\nu }$, while $c\left(
T_{p}\right) /T_{p}\propto \left( \xi _{0}^{T}\right)
^{-3}=V_{c}^{-1}$, it becomes clear that the critical regime
considered here can be attained in superconductors with
sufficiently small correlation length amplitude only. Here$\left|
t_{p}\right| $ can be tuned into the experimentally accessible
range by increasing the magnetic field, while $c\left(
T_{p}\right) /T_{p}$ is controlled by the small inverse
correlation volume.

Since the transverse correlation lengths which are parallel to the
applied field are limited by the length scale $L_{H}$,
Eq.(\ref{eq29}) is readily extended to the anisotropic case for
magnetic fields applied parallel to the a, b or c-axis:
\begin{equation}
\left| t_{p}\right| =\left( \frac{\xi _{0a}^{T}}{L_{H_{a}}}\right)
^{1/\nu },\left| t_{p}\right| =\left( \frac{\xi
_{0b}^{T}}{L_{H_{b}}}\right) ^{1/\nu },\left| t_{p}\right| =\left(
\frac{\xi _{0c}^{T}}{L_{H_{a}}}\right) ^{1/\nu }  \label{eq31}
\end{equation}
where
\begin{equation}
L_{H_{a}}^{2}=\frac{\Phi _{0}}{aH_{a}},\ L_{H_{b}}^{2}=\frac{\Phi
_{0}}{aH_{b}},\ L_{H_{c}}^{2}=\frac{\Phi _{0}}{aH_{c}}.
\label{eq33}
\end{equation}
The specific heat coefficient at $T_{p}$ adopts then the form
given by Eq.(\ref{eq7}) with the correlation volume appropriate
for the anisotropic case. This behavior applies when the condition
$L_{H}<L$ (Eq.(\ref{eq10})), extended to the anisotropic case,
\begin{equation}
L_{H_{a}}^{2}<L_{b}L_{c},\ L_{H_{b}}^{2}<L_{a}L_{c},\
L_{H_{c}}^{2}<L_{a}L_{b},  \label{eq34}
\end{equation}
is satisfied. $L_{i}$, $i=\left( a,\ b,\ c\right) $ denote the
length scales of the sample inhomogeneities. Accordingly, the
occurrence of a magnetic field induced finite size effect requires
that the magnetic field strength exceeds the lower bounds
\begin{equation}
H_{a}>\frac{\Phi _{0}}{aL_{b}L_{c}},\ H_{b}>\frac{\Phi
_{0}}{aL_{a}L_{c}},\ H_{c}>\frac{\Phi _{0}}{aL_{a}L_{b}}.
\label{eq35}
\end{equation}
Otherwise the finite size effect is due to inhomogeneities. When
the magnetic field induced finite size effect dominates, the
scaling form (\ref {eq21}) of the specific heat coefficient adopts
then with Eq.(\ref{eq31}) or Eq.(\ref{eq28a}) the form,
\begin{equation}
\left( \frac{c\left( t,H_{c}\right) }{T}-B^{-}\right) /\left(
\frac{A^{-}}{\alpha }\left| t\right| ^{-\alpha }\right) =g\left(
y\right) ,\ y=t\left( \frac{\Phi _{0}}{aH_{c}\left( \xi
_{0c}^{T}\right) ^{2}}\right) ^{1/2\nu }. \label{eq36}
\end{equation}
Accordingly, given estimates for $A^{-}/\alpha $, $B^{-}$ and $\xi
_{0c}^{T}$, determined from experimental zero field data or the
magnetic field dependence of $t_{p}$ (Eq.(\ref{eq31})), their
consistency can be checked in terms of this scaling relation.
Consistency is achieved whenever the data for the specific heat
coefficient in a magnetic field, plotted according to
Eq.(\ref{eq36}), falls on a single curve, which is the scaling
function $g\left( y\right) $. At $T_{p}$, where $y=y_{p}=-1$, we
have $g\left( -1\right) =1$ and Eq.(\ref{eq20}) is recovered,
while for $\pm y\rightarrow 0 $ it diverges as $g\left( y\right)
\propto \left| y\right| ^{\alpha }$ for $\alpha <0$.

At the first-order melting transition there is a discontinuity in
the internal energy associated with coexistence of the Abrikosov
lattice and the vortex liquid. This in turn gives rise to a
$\delta $-function peak in the specific heat\cite{hu,sudbo} and to
a singularity in the scaling function $G_{3}^{-}(z)$ of the free
energy density (Eq.(\ref{eq22})) at some universal value $z_{m}$
of its argument. According to Eq.(\ref{eq22}) the melting line
$T_{m}\left( H_{i}\right) $ is then fixed by

\begin{equation}
\left| t_{m}\left( H_{i}\right) \right| =\left( \frac{H_{i}\left(
\xi _{0i}^{T}\right) ^{2}}{\Phi _{0}z_{m}}\right) ^{1/2\nu }.
\label{eq36a}
\end{equation}
On the other hand there is the $T_{p}\left( H_{i}\right) $, where
$\xi _{i}^{T}\left( H_{i},T\right) $ attains the limiting magnetic
length scale $L_{H_{i}}$. Using Eqs.(\ref{eq31}) and (\ref{eq33})
we find
\begin{equation}
\left| t_{p}\left( H_{i}\right) \right| =\left( \frac{aH_{i}\left(
\xi _{0i}^{T}\right) ^{2}}{\Phi _{0}}\right) ^{1/2\nu },
\label{eq36b}
\end{equation}
and the ratio
\begin{equation}
\frac{\left| t_{p}\left( H_{i}\right) \right| }{\left| t_{m}\left(
H_{i}\right) \right| }=\left( az_{m}\right) ^{1/2\nu },
\label{eq36c}
\end{equation}
turns out to be universal. Because a limiting length scale, like
$L_{H}$ $\propto H^{-1/2}$, also leads to a rounding of
first-order transitions\cite {privf}, the vortex melting
transition is expected to broaden with increasing magnetic field.

We we are now prepared to analyze experimental data in terms of
the finite size scaling theory.

\section{Comparison with experiment}

We have seen that measurements of thermodynamic properties in an
applied magnetic field open a rather direct route to trace thermal
fluctuations, to estimate various critical properties, to derive
an upper bounds for the length scales associated with
inhomogeneities, to assess the relevance of critical fluctuations
and to probe the interplay of the finite size effects stemming
from sample inhomogeneities and the applied magnetic field. In
particular, the relevance of thermal fluctuations is established
whenever there is a remnant of the zero field singularity at
$T_{p}$, shifting to a lower value with increasing magnetic field
strength. In this case condition (\ref{eq31}) is satisfied and the
field dependence of the relative shift $t_{p}$ should evolve
according to Eq.(\ref{eq36b}). However, due to unavoidable
inhomogeneities the available experimental data do not extend to
the asymptotic critical regime, as required to deduce estimates
for the critical exponents. Since type II superconductors fall in
the experimentally accessible critical regime into the 3D-XY
universality class, we take the critical exponents and critical
amplitude combinations, listed in Eq.(\ref {eq19}), for granted.

\begin{figure}[tbp]
\centering
\includegraphics[totalheight=7cm]{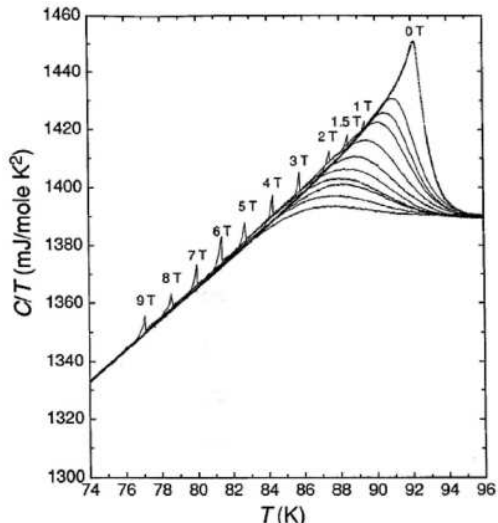}
\caption{Total specific heat of an untwined
YBa$_{2}$Cu$_{3}$O$_{7-\delta }$ single crystal with $T_{c}=91.87$
K for $H_{c}=H\Vert c$. The numbers on top of the spike indicate
the strength of the applied magnetic field and the location of the
melting transition. Taken from Schilling \emph{et
al.}\protect\cite {schilling}} \label{fig1}
\end{figure}

\begin{figure}[tbp]
\centering
\includegraphics[totalheight=7cm]{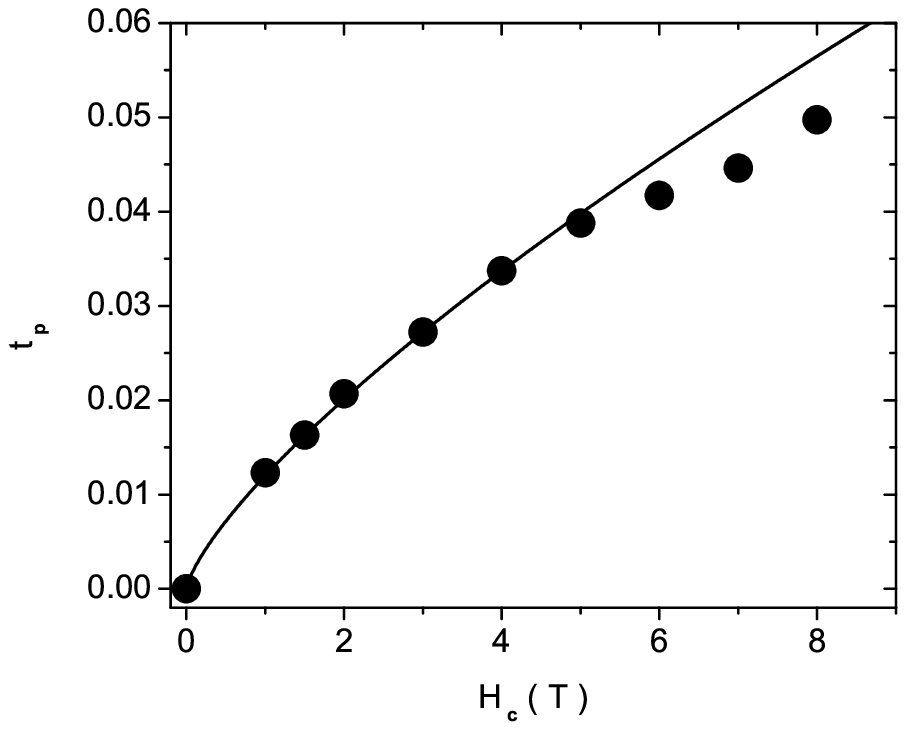}
\includegraphics[totalheight=7cm]{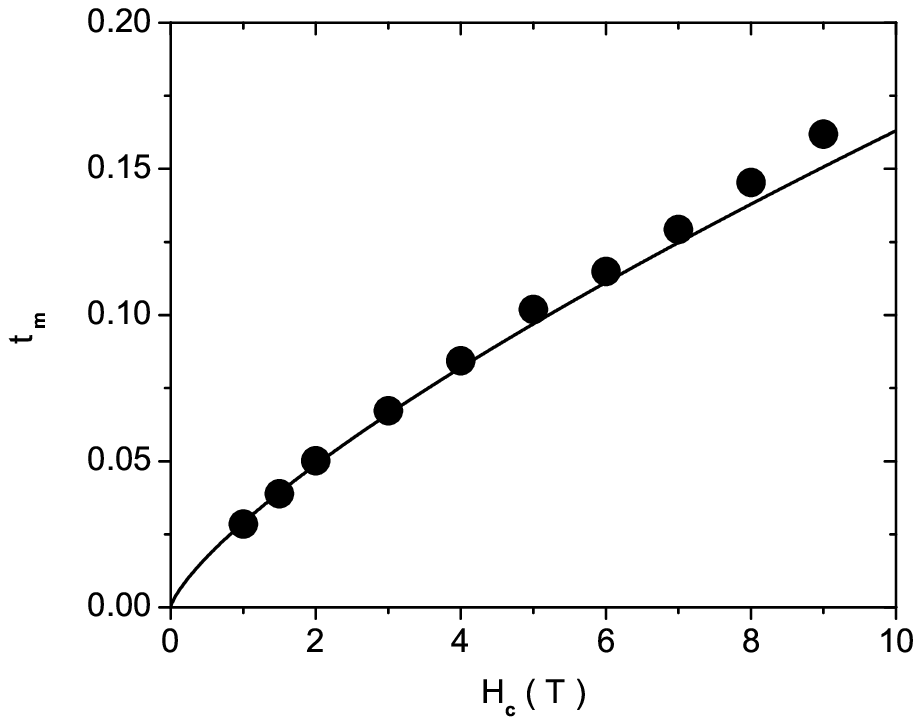}
\caption{(a)$\left| t_{p}\right| $ versus $H_{c}$ for the data
shown in Fig.\ref{fig1}. The solid line is Eq.(\ref{eq37}) with
$\nu $ listed in Eq.(\ref {eq19}). (b) Melting line $T_{m}\left(
H_{c}\right) $ in terms of $\left| t_{m}\right| $versus $H_{c}$
for the data shown in Fig.\ref{fig1}. The solid line is
Eq.(\ref{eq38b}) } \label{fig2}
\end{figure}

In Fig.\ref{fig1} we displayed the data of Schilling \emph{et
al.}\cite {schilling} for the temperature dependence of the heat
coefficient of a YBa$_{2}$Cu$_{3}$O$_{7-\delta }$ single crystal
with $T_{c}=91.87$ K at various magnetic fields applied parallel
to the c-axis ($H_{c}$). As a remnant of the zero field
singularity, there is for fixed field strength a broad peak
adopting its maximum at $T_{p}$ which is located below $T_{c}$. As
$T_{p}$ approaches $T_{c}$, the peak becomes sharper with
decreasing $H$ and evolves smoothly to the zero-field singularity,
smeared by the inhomogeneity induced finite size effect. The spike
at $T_{m}<T_{p}$ is due to the melting transition\cite{schilling}.
Since $T_{p}$ increases systematically with increasing field,
condition (\ref{eq34}) is satisfied and there is a magnetic field
induced finite size effect. Accordingly Eq.(\ref{eq36b}) for the
field dependence of the relative shift should apply as long as 3D
critical fluctuations dominate. In Fig.\ref{fig2} we displayed
$t_{p}$ versus $H$ for the data shown in Fig.\ref{fig1}. The solid
line is Eq.(\ref {eq36b}) with
\begin{equation}
\left| t_{p}\right| =0.012\ H_{c}^{1/2\nu },  \label{eq37}
\end{equation}
$H$ in T and $\nu $ given by Eq.(\ref{eq19}). In the low field
regime, where this asymptotic form applies it describes the data
remarkably well, while for $\left| t_{p}\right| \gtrsim 0.04$
deviations pointing to corrections to scaling become manifest. To
fix the universal constant $a$ entering Eq.(\ref {eq24a}) we use
for $\xi _{0c}^{T}$ the estimate $\xi _{0c}^{T}=\sqrt{\xi
_{a0}^{-}\xi _{b0}^{-}}=\sqrt{12.4\cdot 14.7}=13.5$A, derived from
magnetic torque measurements\cite{book}, Eqs.(\ref{eq36b}) and
(\ref{eq37}) then yield for the universal constant $a$ entering
the limiting length scales the estimate

\begin{equation}
a\approx 3.12.  \label{eq38}
\end{equation}
In $^{4}$He this value leads with $L_{\Omega }=\sqrt{h/\left(
2am_{4}\Omega \right) }$ and $\Omega =\left( 2\pi \right)
^{-1}$s$^{-1}$ to $L_{\Omega }\approx 5$ $10^{-2}$cm, which
compares reasonably well with $5$ $10^{-2}$cm, the estimate
derived by Haussmann\cite{hauss}. Since at the lowest attained
magnetic field, $H_{c}=1$T, (see Fig.\ref{fig1}) a shift from
$T_{c} $ to $T_{p}$ occurs, Eq.(\ref{eq36b}) yields for the
inhomogeneities in the ab--plane the lower bound $L_{ab}>257A$,
which is close to the estimates derived from the finite size
effect in the zero field specific heat coefficient of untwined
YBa$_{2}$Cu$_{3}$O$_{7-\delta }$ single crystals
\cite{book,housten}.

The existence of the first order melting transition of the vortex
lattice, seen in Fig.\ref{fig1} in terms of the spikes with
numbers on the top, implies that the scaling function
$G_{3}^{-}(z)$ in the free energy density (Eq.(\ref{eq22}))
exhibits at some universal value $z_{m}$ of its argument a smeared
singularity. For a magnetic field applied along the c-axis the
melting line is then given by Eq.(\ref{eq36b}). In Fig.\ref{fig2}b
we compare this prediction with the experimental melting line,
derived from the data shown in Fig.\ref{fig1}, in terms of the
solid line, which is
\begin{equation}
\left| t_{m}\right| =0.029\ H_{c}^{1/2\nu }.  \label{eq38b}
\end{equation}
Together with $\xi _{0c}^{T}=13.5$A, the value used to estimate
$a$ (Eq.(\ref {Eq38})), we obtain for the universal value of
$z_{m}$ the estimate
\begin{equation}
z_{m}\approx 0.1.  \label{eq38c}
\end{equation}
Since a limiting length scale, like $L_{H}$ $\propto H^{-1/2}$,
leads also to a rounding of first-order transitions\cite{privf},
the vortex melting transition is expected to become broader with
increasing magnetic field. This behavior appears to be confirmed
by the extended study of the vortex melting transition in
YBa$_{2}$Cu$_{3}$O$_{7-\delta }$ single crystals of Roulin
\emph{et al.}\cite{roulin}.

\begin{figure}[tbp]
\centering
\includegraphics[totalheight=7cm]{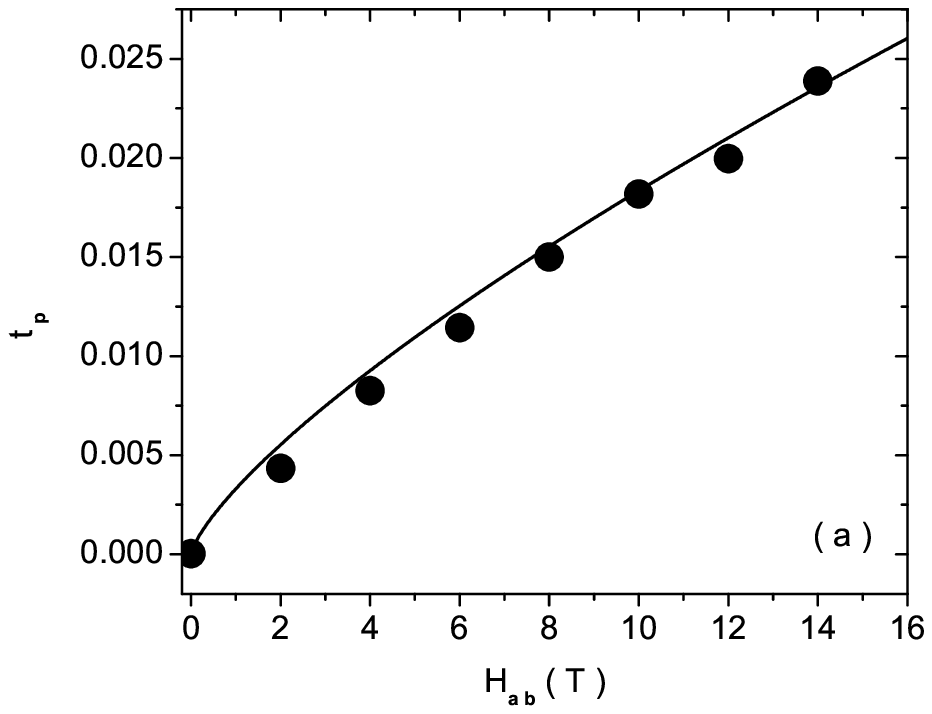}
\includegraphics[totalheight=7cm]{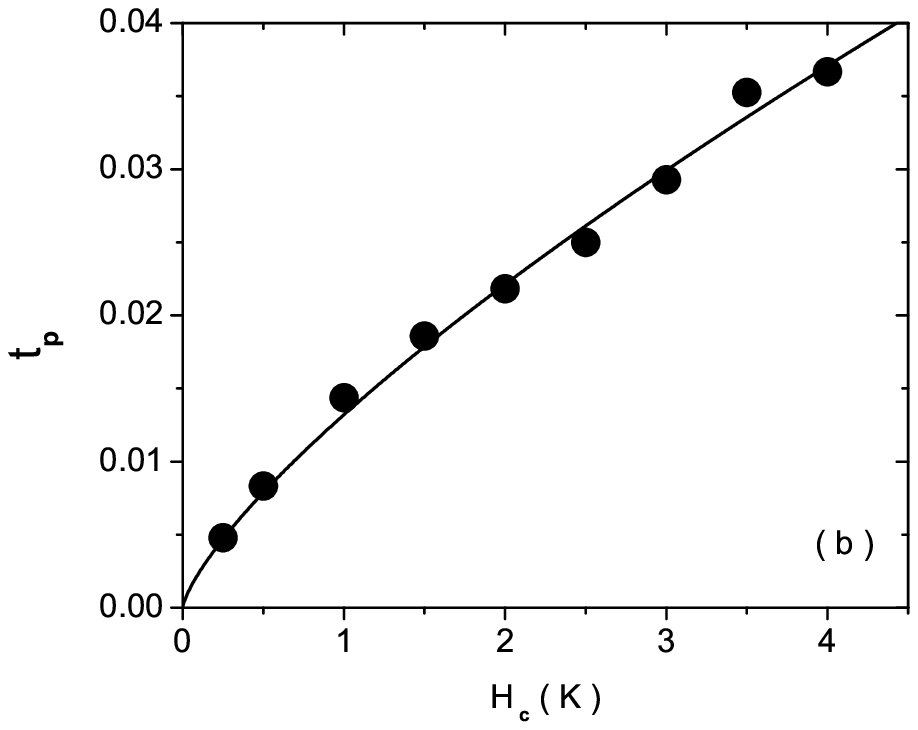}
\caption{$\left| t_{p}\right| $ versus $H$ for
YBa$_{2}$Cu$_{3}$O$_{6.9}$ with $T_{c}=92.56$ K derived from the
data of Roulin \emph{et al.}\protect\cite{roulin}. (a) $\left|
t_{p}\right| $ versus $H_{ab}$; (b) $\left| t_{p}\right| $ versus
$H_{c}$. The solid lines are Eqs.(\ref{eq39}) with $\nu $ listed
in Eq.(\ref{eq19}).} \label{fig3}
\end{figure}

To explore variations between data and single crystals of
different provenance we show in Fig.\ref{fig3} t$_{p}$ versus
$H_{ab}$ and $H_{c}$ for YBa$_{2}$Cu$_{3}$O$_{6.9}$ single crystal
with $T_{c}=92.56$K derived from the data of Roulin \emph{et
al.}\cite{roulin}. The solid lines are
\begin{equation}
\left| t_{p}\left( H_{ab}\right) \right| =0.0033\ H_{ab}^{1/2\nu
},\ \left| t_{p}\left( H_{c}\right) \right| =0.0132\ H_{c}^{1/2\nu
},  \label{eq39}
\end{equation}
corresponding to Eq.(\ref{eq36b}) with $H$ in T and $\nu $ given
by Eq.(\ref {eq19}). They yield with $a=3.12$ (Eq.(\ref{eq38}))
the estimates
\begin{equation}
\xi _{0ab}^{T}=5.7\text{A},\ \xi _{0c}^{T}=14.4\text{A,
}V_{c}^{-}=\left( \xi _{0ab}^{T}\right) ^{2}\xi
_{0c}^{T}=469\text{A}^{3}  \label{eq40}
\end{equation}
Since at the lowest attained fields $H_{c}=0.25$T and $H_{ab}=2$T
a shift from $T_{c}$ to $T_{p}$ is still present, we derive from
Eq.(\ref{eq35}) for the length scales of the sample
inhomogeneities the lower bounds
\begin{equation}
L_{ab}>515\text{A},\ \ \sqrt{L_{ab}L_{c}}>182\text{A}.
\label{eq40a}
\end{equation}

The reported specific heat data also revealed melting lines
consistent with the scaling form (\ref{eq36b}), namely

\begin{equation}
\left| t_{m}\left( H_{ab}\right) \right| =0.008\ H_{ab}^{1/2\nu
},\ \left| t_{m}\left( H_{c}\right) \right| =0.032\ H_{c}^{1/2\nu
},  \label{eq40b}
\end{equation}
with $\nu =2/3$, which is close to the 3D-XY value
(Eq.(\ref{eq19})). Together with the expressions for the
$t_{p}\left( H_{i}\right) $ lines (Eq.(\ref{eq39})) we obtain for
the universal ratio $\left| t_{p}\left( H_{i}\right) \right|
/\left| t_{m}\left( H_{i}\right) \right| =\left( az_{m}\right)
^{1/2\nu }$ (Eq.\ref{eq36c})) the estimates, $\left| t_{p}\left(
H_{ab}\right) \right| /\left| t_{m}\left( H_{ab}\right) \right|
=0.41$ and $\left| t_{p}\left( H_{c}\right) \right| /\left|
t_{m}\left( H_{c}\right) \right| =0.42$, which agree with $\left(
az_{m}\right) ^{1/2\nu }=\left( 3.12\ 0.1\right) ^{3/4}$, derived
from the independent estimates for $a$ (Eq.(\ref{eq38}) and
$z_{m}$ (Eq.(\ref{eq38c})).

\begin{figure}[tbp]
\centering
\includegraphics[totalheight=7cm]{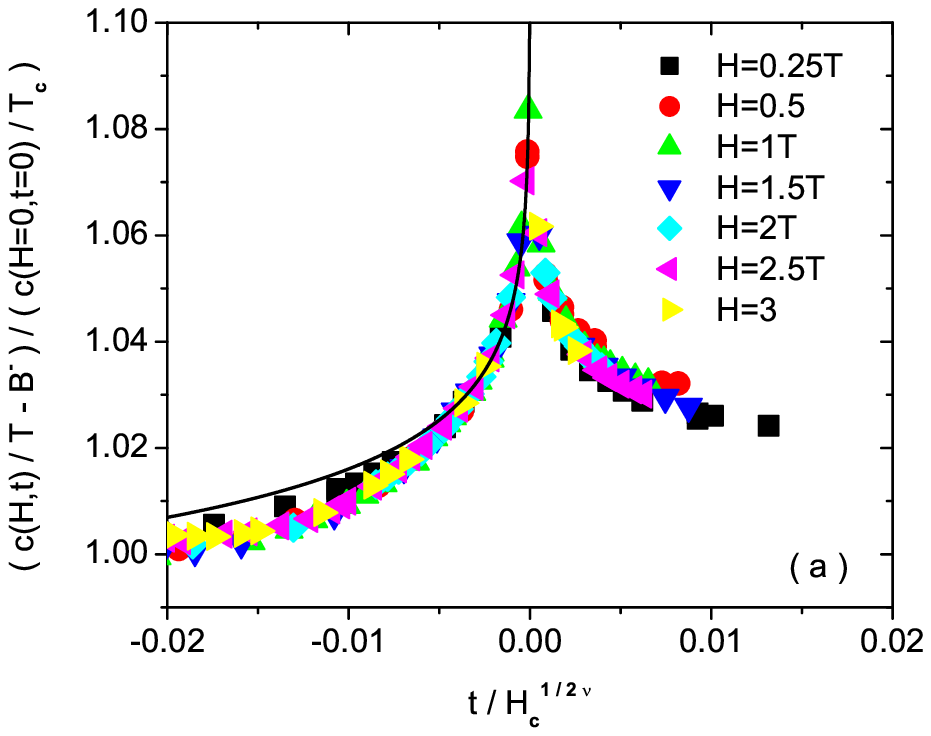}
\includegraphics[totalheight=7cm]{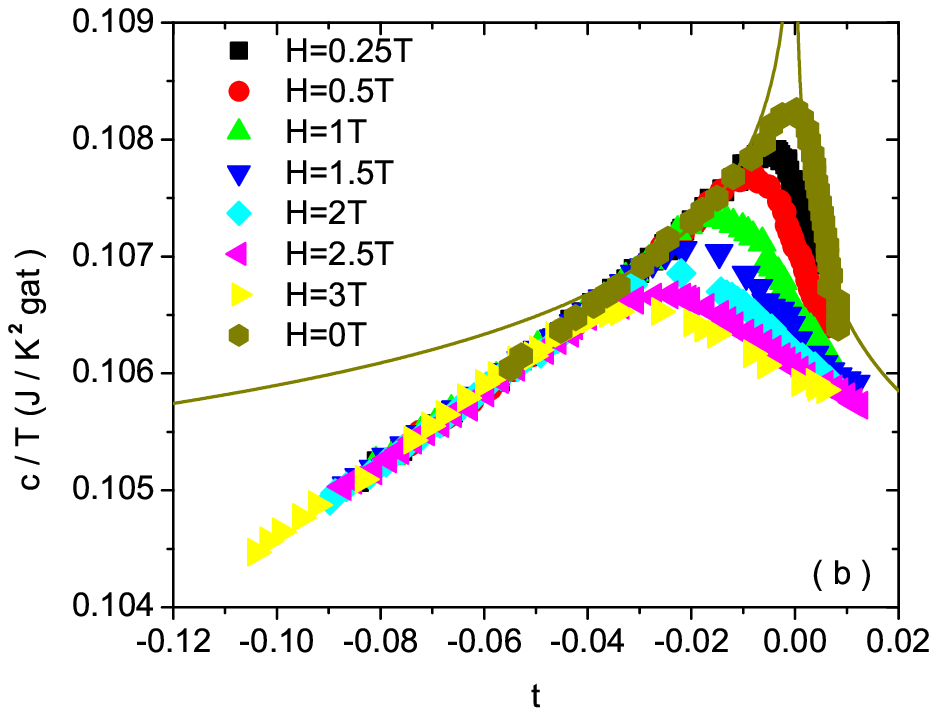}
\caption{ (a):\ $\left( c\left( T,H_{c}\right) /T-B^{-}\right)
/\left( A^{-}\left| t\right| ^{-\alpha }/\alpha \right) $ versus
$t/H_{c}^{1/2\nu }$ for YBa$_{2}$Cu$_{3}$O$_{6.9}$ with
$T_{c}=92.56$ K derived from the data of Roulin \emph{et al.}
\protect\cite{roulin}, with $\alpha $ and $\nu $ given by
Eq.(\ref{eq19}), $\left( c\left( H,t\right) /T-B^{-}\right) $ in
J/K$^{2}$gat and $H$ in T. The values for $A^{-}/\alpha $ and
$B^{-}$ are listed in Eq.(\ref{eq42}). The solid line is
Eq.(\ref{eq41}); (b): \ Specific heat coefficient $c/T$ versus $t$
of YBa$_{2}$Cu$_{3}$O$_{6.9}$ with $T_{c}=92.56$ K at various
magnetic fields derived from the data of Roulin \emph{et
al.}\protect\cite {roulin}. The solid lines are Eq.(\ref{eq16})
with $\alpha $ and $A^{+}/A^{-} $ given by Eq.(\ref{eq19}) and the
parameters listed in Eq.(\ref {eq42}). The deviations of the zero
field data from these lines reveal around $t=0$ the finite size
effect due to inhomogeneities.} \label{fig4}
\end{figure}

As the data of Roulin \emph{et al.}\cite{roulin} are rather dense
and extend close to zero field criticality, a more detailed finite
size scaling analysis appears to be feasible. Using
Eq.(\ref{eq36}), it is seen from Fig.\ref{fig4}a that data
collapse is obtained over a rather wide range of the scaling
variable. Indeed, finite size scaling requires the data to
collapse on the universal curve $g\left( y\right) $ with $\
y=t\left( \Phi _{0}/\left( aH_{c}\left( \xi _{0c}^{T}\right)
^{2}\right) \right) ^{1/2\nu }$. Furthermore, scaling also
requires that at $T_{p}$, where $y=y_{p}=-1$, and for $\pm
y\rightarrow 0$, $g\left( -1\right) =1$ and  $g\left( y\right)
\propto \left| y\right| ^{-\left| \alpha \right| }$ for $\alpha
<0$ should hold, respectively. Using $\xi _{0c}^{T}=14.4$A
(Eq.(\ref{eq40})), it is seen from Fig.\ref{fig4}a that at
$y_{p}$, where $\left| t\right| /H_{c}^{1/2}=-0.013$, the scaling
function is close to $1$, as required. In addition, the solid line
which is
\begin{equation}
\frac{c\left( T,H_{c}\right) /T-B^{-}}{\left( A^{-}\left| t\right|
^{-\alpha }/\alpha \right) }=0.957\left(
\frac{-t}{H_{c}^{1/2}}\right) ^{\alpha }, \label{eq41}
\end{equation}
also confirms the divergence of the scaling function at $\pm y=0$.
To qualify the remarkable collapse of the data we note that the
estimates for $\widetilde{A}^{-}=A^{-}/\alpha $ and $B^{-}$ have
been derived from the zero field data displayed in
Fig.\ref{fig4}b.\ The solid line for $t<0$ is Eq.(\ref{eq16}) with
\begin{equation}
\frac{A^{-}}{\alpha
}=\widetilde{A}^{-}=-0.073\text{(J/K}^{2}\text{gat)},\ \
B^{-}=0.179\text{(J/K}^{2}\text{gat),
}B^{+}=0.181\text{(J/K}^{2}\text{gat), }T_{c}=92.56\text{K}
\label{eq42}
\end{equation}
and the critical exponent $\alpha $ listed in Eq.(\ref{eq19}), to
indicate the inhomogeneity induced deviations from the leading
zero field critical behavior of \ perfect
YBa$_{2}$Cu$_{3}$O$_{6.9}$. Since the data collapse onto the
universal finite scaling function has been achieved without any
arbitrary fitting parameter, the consistency with the magnetic
field induced finite size scenario is well confirmed. Furthermore,
Fig.\ref{fig4}b shows how the broad anomaly in the specific heat
coefficient sharpens and that the maximum height at $T_{p}\left(
H_{c}\right) $ increases, evolving smoothly to the zero field
peak, rounded by inhomogeneities.

To estimate the correlation volume from the specific heat
coefficient in terms of the universal relation Eq.(\ref{eq18}) we
note that $A^{-}=T_{c}\alpha $ $\widetilde{A}^{-}$,
$A^{-}$(cm$^{-3}$)$=\left( 10^{4}/k_{B}/V_{gat}\right)
A^{-}$(mJ/K/cm$^{3}$) and $V_{gat}=8$cm$^{3}$ give $A^{-}=7.96\
10^{-4}$A$^{-3}$, yielding with $R^{-}=0.815$ (Eq.(\ref {eq19}))
the correlation volume $V_{c}^{-}\approx 680$A$^{3}$, which is
reasonably close to the value derived from the magnetic field
induced finite size effect (Eq.(\ref{eq40})). As in zero field the
correlation volume $V_{c}^{-}\left| t_{p}\right| ^{-3\nu }$ cannot
grow beyond the volume $V_{i}$ of the inhomogeneities, this
limiting volume scale is obtained from
\begin{equation}
V_{i}=V_{c}^{-}\left| t_{p}\right| ^{-3\nu }.  \label{eq43}
\end{equation}
and $t_{p}$ evaluated in zero field. With $V_{c}^{-}=469$A$^{3}$
(Eq.(\ref {eq40})) and $\left| t_{p}\right| =0.0025$ taken from
Fig.\ref{fig4}b, we obtain the estimates $V_{i}=7.5\
10^{7}$A$^{3}$ and $V_{i}^{1/3}=422$A which is consistent with the
aforementioned lower bounds listed in Eq.(\ref{eq40a}).

\begin{figure}[tbp]
\centering
\includegraphics[totalheight=7cm]{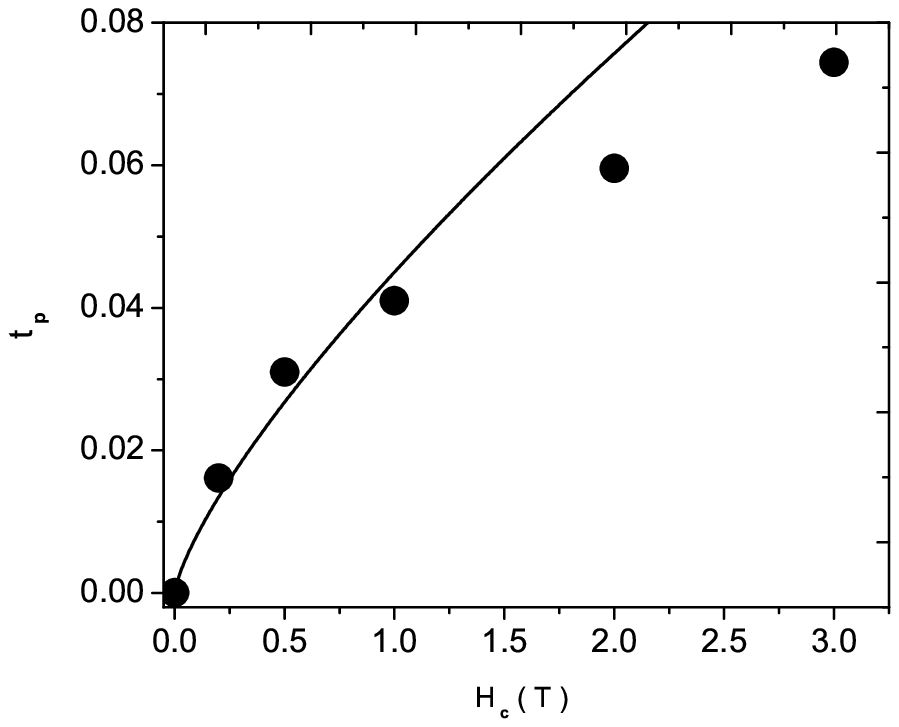}
\includegraphics[totalheight=7cm]{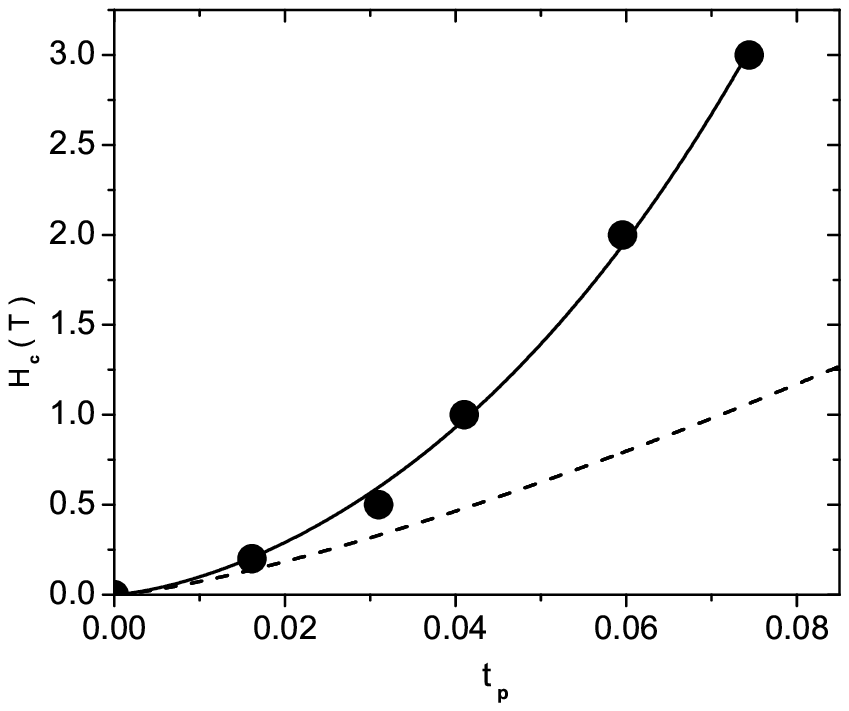}
\caption{(a) $\left| t_{p}\right| $ versus $H_{c}$ for
YBa$_{2}$Cu$_{3}$O$_{6.6}$ with $T_{c}=64.2$K derived from the
data of Junod \emph{et al.}\protect\cite {junod}. The solid line
is Eq.(\ref{eq41}) with $\nu $ listed in Eq.(\ref {eq19}). (b)
$H_{c}$ versus \ $\left| t_{p}\right| $ of the same data. The
solid curve is Eq.(\ref{eq48}) which takes the correction to the
leading scaling term, $H_{c}=33.97\left| t_{p}\right| ^{2\nu }$,
displayed by the dashed line, into account.} \label{fig5}
\end{figure}

Noting that the critical amplitudes, the anisotropy $\gamma $
$=\left( \xi _{ab0}^{T}/\xi _{c0}^{T}\right) ^{1/2}$ and the
length scale of the inhomogeneities depend on the dopant
concentration\cite{parks} it is instructive to analyze the data of
Junod \emph{et al.} \cite{junod} for underdoped
YBa$_{2}$Cu$_{3}$O$_{6.6}$ with $T_{c}=64.2$K and $\gamma \approx
20$\cite{janossy}. In Fig.\ref{fig5} we displayed the data for
$\left| t_{p}\right| $ versus $H_{c}$ together with
\begin{equation}
\left| t_{p}\left( H_{c}\right) \right| =0.045\ H_{c}^{1/2\nu },
\label{eq44}
\end{equation}
corresponding to Eq.(\ref{eq36b}) with $H$ in T and $\nu $ given
by Eq.(\ref {eq19}). It yields with $a=3.12$ (Eq.(\ref{eq38})) the
estimate
\begin{equation}
\xi _{0c}^{T}=32.6\text{A,}  \label{eq44a}
\end{equation}
compared to $\ \xi _{0c}^{T}=14.4$ A (Eq.(\ref{eq40})) obtained
close to optimum doping. Thus, the critical amplitude of \ the
correlation length $\xi _{0c}^{T}$ increases in the underdoped
regime, reflecting the approach to the 2D quantum superconductor
to insulator transition in the underdoped limit, where $\gamma $
tends to infinity\cite{book,parks}. Typically, the leading scaling
behavior becomes valid when $\left| t\right| $ becomes small and
$H$ is small. However, one needs to know: How small is small
enough? As there are corrections to the leading scaling behavior
we know, when $\left| t\right| $ is no longer small, the
temperature dependence of the transverse correlation length scales
as\cite{privman,peliasetto,book}
\begin{equation}
\xi _{c}^{T}\left( t\right) =\xi _{0c}^{T}\left| t\right| ^{-\nu
}\left( 1+a_{\xi _{c}^{T}}\left| t\right| ^{\omega \nu }+..\right)
\label{eq45}
\end{equation}
where $\omega \nu $ is the correction to scaling exponent,
adopting in the 3D-XY universality class the
value\cite{peliasetto}
\begin{equation}
\omega \nu =0.53.  \label{eq46}
\end{equation}
The dimensionless correction amplitude $a_{\xi ^{T}}\ $will in
general become larger near a crossover. As
YBa$_{2}$Cu$_{3}$O$_{7-\delta }$ undergoes with reduced oxygen
concentration a 3D-2D crossover, $a_{\xi _{c}^{T}}$ is expected to
increase, while the critical regime where the leading term
dominates shrinks. To substantiate this expectation and to
illustrate the effect of the corrections to scaling, we set \ $\xi
_{c}^{T}\left( T,H\right) =L_{H_{c}}$, to obtain with
Eqs.(\ref{eq31}) and (\ref{eq33}) for $H_{c}\left( t_{p}\right) $
the expression
\begin{equation}
H_{c}\left( t_{p}\right) =\frac{\Phi _{0}}{a\left( \xi
_{0c}^{T}\right) ^{2}}\left| t_{p}\right| ^{2\nu }\left( 1+a_{\xi
_{c}^{T}}\left| t_{p}\right| ^{\omega \nu }\right) ^{-2}.
\label{eq47}
\end{equation}
A fit to the experimental data shown in Fig.\ref{fig5}b yields
\begin{equation}
H_{c}=33.97\left| t_{p}\right| ^{2\nu }\left( 1-1.61\left|
t_{p}\right| ^{\omega \nu }\right) ^{-2},  \label{eq48}
\end{equation}
with $\nu $ and $\omega \nu $ given by Eqs.(\ref{eq19}) and
(\ref{eq46}), respectively. This gives $\xi _{0c}^{T}=40.86$A
which is considerably larger than $\xi _{0c}^{T}=32.6$A, the value
derived without the correction to scaling. On the contrary, a fit
to the data displayed in Fig.\ref{fig3}b yields
$H_{c}=319.66\left| t_{p}\right| ^{2\nu }\left( 1-0.014\left|
t_{p}\right| ^{\omega \nu }\right) ^{-2}$, revealing that close to
optimum doping the correction is up to $\left| t_{p}\right| =0.04$
negligibly small, while in the underdoped sample it is significant
down to $\left| t_{p}\right| =0.02$. This confirms the expectation
whereupon the 3D-2D crossover in the underdoped regime enhances
the correction to scaling and reduces the critical regime where
the leading scaling term dominates. From the data of Junod
\emph{et al.} \cite{junod} we also deduce that at the lowest
attained field, $H_{c}=0.2$T, a shift from $T_{c}$ to $T_{p}$
occurs. This yields with Eq.(\ref{eq35}) the lower bound
\begin{equation}
\sqrt{L_{a}L_{b}}>576\text{A}  \label{eq48a}
\end{equation}
compared to $515$ A, the value close to optimum doping (Eq.(\ref{eq40a})).

\begin{figure}[tbp]
\centering
\includegraphics[totalheight=7cm]{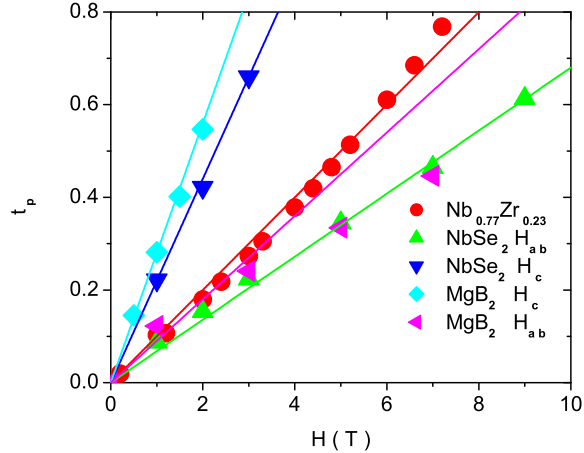}
\caption{$\left| t_{p}\right| $ versus $H$ for
Nb$_{77}$Zr$_{23}$\protect\cite {mirmelstein},
2H-NbSe$_{2}$\protect\cite{sanchez} and
MgB$_{2}$\protect\cite{lyard} derived from the respective
references. The straight lines are Eq.(\ref{eq49}) with the
parameters listed in Table I.} \label{fig6}
\end{figure}

Next we turn to Nb$_{77}$Zr$_{23}$\cite{mirmelstein},
2H-NbSe$_{2}$\cite {sanchez} and MgB$_{2}$\cite{lyard}, type II
superconductors supposed to have comparative large correlation
volumes. In Fig.\ref{fig6} we displayed $\left| t_{p}\right| $
versus $H$ derived from the respective experimental data. In
contrast to the corresponding plot for nearly optimally
YBa$_{2}$Cu$_{3}$O$_{7-\delta }$ (Fig.\ref{fig3}), the data points
to a linear relationship. Consequently, the critical regime, where
$\left| t_{p}\right| $ $\propto $ $H^{1/2\nu \text{ }}$with $\nu
\approx 2/3$ holds (Eq.(\ref{eq36b})), is not attained.
Nevertheless, there is clear evidence for a magnetic field induced
finite size effect, because $T_{p}$ shifts monotonically to a
lower value with increasing magnetic field. Since the data points
to an effective critical exponent $\nu \approx 1/2$ which applies
over an unexpectedly extended range, we use Eq.(\ref{eq36b}) with
$\nu =1/2$ to derive estimates for the amplitude of the respective
transverse correlation lengths and the correlation volume in terms
of
\begin{equation}
\left| t_{p}\right| =b_{i}H_{i}=\frac{\left( \xi _{0i}^{T}\right)
^{2}}{L_{H_{i}}^{2}}=\frac{aH_{i}\left( \xi _{0i}^{T}\right)
^{2}}{\Phi _{0}}, \label{eq49}
\end{equation}
with $a=3.12$ (Eq.(\ref{eq38}). The respective straight lines in
Fig.\ref {fig6} are this relation with the parameters listed in
Table I. For comparison we included the corresponding parameters
for YBa$_{2}$Cu$_{3}$O$_{6.9}$ and YBa$_{2}$Cu$_{3}$O$_{6.6}$
where the critical regime is attained. It is evident that
Nb$_{77}$Zr$_{23}$, 2H-NbSe and MgB$_{2}$ are type II
superconductors with comparatively large correlation lengths.
Compared to YBa$_{2}$Cu$_{3}$O$_{6.9}$ and
YBa$_{2}$Cu$_{3}$O$_{6.6}$ the correlation volume is 3 orders of
magnitude larger, which renders the amplitude of the specific heat
singularity very weak (see Eq.(\ref{eq16}). Nevertheless, the
unambiguous evidence for the magnetic field induced finite size
effect reveals that fluctuations, even though not critical, are at
work. For this reason there is no critical line $T_{c2}\left(
H\right) $ of continuous phase transitions, but a line
$T_{p}\left( H\right) $ where the specific heat peak, broadening
and decreasing with increasing field, adopts its maximum value and
the correlation length attains the limiting magnetic length scale
$L_{H}$. Because the fluctuations are also subject to the finite
size effect arising from inhomogeneities with length scale $L$,
the magnetic finite size effect is observable as long as
$L>L_{H}$. The resulting lower bounds for the length scale of
inhomogeneities affecting the thermodynamic properties are also
included in Table I. Noting that these bounds stem from studies
where no attempt was made to explore the low field behavior, it is
conceivable that the actual length scale of the inhomogeneities is
much larger. To our best knowledge, the only absolute reference
stems from the finite size scaling analysis of the zero field
specific heat data of nearly optimally doped
YBa$_{2}$Cu$_{3}$O$_{7-\delta }$, where $L$ was found to range
from $290$ to $419$A \cite{housten,book}. Interestingly enough,
the largest bound found here applies to the cubic superconducting
alloy Nb$_{77}$Zr$_{23}$. In any case, the lower bounds ranging
from $182$ to $814$A raise serious doubts on the relevance of the
nanoscale spatial variations in the electronic characteristics
observed in underdoped Bi-2212 with scanning tunnelling
microscopy\cite {liu,chang,cren,lang}.

\bigskip

\begin{tabular}{|l|l|l|l|l|l|l|l|l|}
\hline
& $T_{c}$(K) & $b,b_{ab}$ & $b_{c}$ & $\xi _{0}^{T},\ \xi _{0ab}^{T}$(A) & $%
\xi _{0c}^{T}$(A) & V$_{c}^{-}$(A$^{3}$) & $L,\ L_{ab}$(A) & $\sqrt{%
L_{ab}L_{c}}$(A) \\ \hline
Nb$_{77}$Zr$_{23}$ & $10.79$ & $0.1$ & $-$ & $55.1$ & $-$ & $1.7\ 10^{5}$ & $%
>814$ & $-$ \\ \hline
2H-NbSe$_{2}$ & $7.1$ & $0.068$ & $0.22$ & $42.91$ & $93.87$ &
$1.7\ 10^{5}$ & $>258$ & $>258$ \\ \hline
MgB$_{2}$ & $35$ & $0.09$ & $0.28$ & $51.73$ & $110.24$ & $2.9\ 10^{5}$ & $%
>364$ & $>258$ \\ \hline
YBa$_{2}$Cu$_{3}$O$_{6.95}$ & $92.56$ & $0.0033$ & $0.0132$ &
$5.7$ & $14.4$ & $4.7\ 10^{2}$ & $>515$ & $>182$ \\ \hline
YBa$_{2}$Cu$_{3}$O$_{6.6}$ & $64.2$ & $-$ & $0.045$ & $-$ & $32.6$ & $-$ & $%
>576$ & $-$ \\ \hline
\end{tabular}

\bigskip

Table I: Summary of the estimates for the transverse correlation
length amplitudes, correlation volume $V_{c}=\left( \xi
_{0ab}^{T}\right) ^{2}\xi _{0c}^{T}$, lower bounds for the length
scales\ $L_{ab}$ and $\sqrt{L_{ab}L_{c}}$ of inhomogeneities,
derived from the magnetic field induced finite size effect. For
Nb$_{77}$Zr$_{23}$, 2H-NbSe$_{2}$\ and MgB$_{2}$ we used
Eq.(\ref{eq49}), describing the intermediate critical behavior
displayed in Fig.\ref{fig6}. Using the fact that at the lowest
attained magnetic fields, $H=0.1$T (Nb$_{77}$Zr$_{23}$)
\cite{mirmelstein}, $H_{ab}=H_{c}=1$T (2H-NbSe$_{2}$)
\cite{sanchez} and $H_{ab}=1$T, $H_{c}=0.5$T
(MgB$_{2}$)\cite{lyard}, a shift from $T_{c}$ to $T_{p}$ was
observed, we obtain with Eq.(\ref{eq35}) the quoted lower bounds
for the length scale of the inhomogeneities which affect the
thermodynamic properties. For YBa$_{2}$Cu$_{3}$O$_{7-\delta }$ the
estimates stem from the critical behavior shown in Figs.\ref{fig3}
and \ref{fig5}, quoted in Eqs.(\ref{eq39}), (\ref{eq40}),
(\ref{eq44}) and (\ref{eq44a}), and the lower bounds are
Eqs.(\ref{eq40a}) and (\ref{eq48a}).

\bigskip

\bigskip

\begin{figure}[tbp]
\centering
\includegraphics[totalheight=7cm]{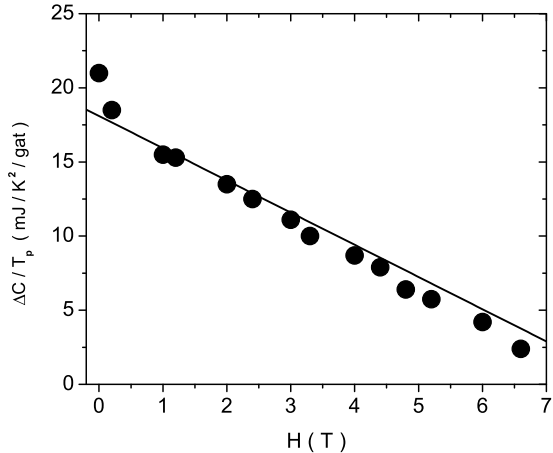}
\caption{$\Delta c\left( H,T_{p}\right) /T_{p}$ versus $H$ for
Nb$_{77}$Zr$_{23}$. Experimental data taken from Mirmelstein
\emph{et al}.\protect\cite{mirmelstein}. The solid line is
Eq.(\ref{eq2}) rewritten in the form (\ref {eq50}) and the
parameters listed in Eq.(\ref{eq51}).} \label{fig7}
\end{figure}

The occurrence of fluctuations in type II superconductors with
comparatively large correlation lengths also implies a rounding of
the specific heat peak in zero field, where inhomogeneities set
the limiting length scale. Such a rounding was found in the
specific heat data of polycrystalline Mg$^{11}$B$_{2}$, obtained
with high resolution ac calorimetry\cite{park}. Further evidence
for the relevance of fluctuations can be deduced from the specific
heat coefficient at $T_{p}$. In Fig.\ref{fig7} we show the
experimental data for Nb$_{77}$Zr$_{23}$\cite{mirmelstein}. For
comparison we included Eq.(\ref {eq2}) in the form
\begin{equation}
\frac{\Delta c\left( H,T_{p}\right) }{T_{p}}=c\left( 1-bH\right)
\label{eq50}
\end{equation}
with the parameters
\begin{equation}
c=18.1\text{mJ/K}^{2}\text{gat, \ \ }b=0.12\text{T}^{-1}.
\label{eq51}
\end{equation}
Here fluctuations enter via the coefficient $b$ (Eq.(\ref{eq49})).
For $H>0.2 $T the data is consistent with a linear relationship
and the coefficient $b$ derived from the field dependence of
$T_{p}$ (see Table I). The difference between the zero field value
$\Delta c\left( H=0,T_{c}\right) /T_{c}c=21$mJ/K$^{2}$gat and the
limit $\Delta c\left( H\rightarrow 0,T_{p}\right) /T_{p}$
$=c=18.1$mJ/K$^{2}$gat is due to the triangular vortex lattice,
for which the ratio between these coefficients should be
$1.16$\cite{fetter,mirmelstein}.

Although we concentrated hitherto on the specific heat, it should
be kept in mind that the magnetic field induced finite size effect
also affects other thermodynamic and the transport properties,
including the magnetoconductivity. As an example we consider the
linear magentoconductivity of a bulk sample. In the isotropic case
the fluctuation contribution scales in zero field as $\sigma
\propto \left( \xi ^{T}\right) ^{z-1}$\cite{book}, where $z$ is
the dynamic critical exponent. Together with the evidence for
static 3D-XY critical exponents there is mounting evidence for
$z=2$ \cite{book,tshk,vortexgl,osborn} deduced from zero field
data of the fluctuation contribution to the conductivity. In the
presence of a magnetic field we obtain from Eq.(\ref{eq28a}) the
scaling form
\begin{equation}
\sigma \propto \left( \xi ^{T}\right) ^{z-1}F(y),\text{ \
}y=\frac{aH\left( \xi ^{T}\right) ^{2}}{\Phi _{0}}=\left(
\frac{\xi ^{T}}{L_{H}}\right) ^{2} \label{eq52}
\end{equation}
where $F(y)$ is a universal scaling function of its argument Since
in an applied field the correlation length cannot grow beyond
$L_{H}$, there is at $T_{p}$ the residual conductivity
\begin{equation}
\sigma \left( H_{p}\right) \propto \left( \frac{\Phi
_{0}}{aH_{p}}\right) ^{\left( z-1\right) /2}F(1),  \label{eq53}
\end{equation}
where, using Eq.(\ref{eq24a}),
\begin{equation}
H_{p}\left( T\right) =\frac{\Phi _{0}}{a\xi _{0}^{2}}t^{2\nu
}=\frac{\Phi _{0}}{a\xi _{0}^{2}}\left( 1-\frac{T}{T_{c}}\right)
^{2\nu },  \label{eq54}
\end{equation}
This implies that type II superconductors in a magnetic field do
not undergo a phase transition to a state with zero resistance. In
particular there is no resistive upper critical field
$H_{c2}\left( T\right) $. It is replaced by the line $H_{p}\left(
T\right) $ where the resistivity attains the residual value $\rho
\left( H_{p}\right) =1/\sigma \left( H_{p}\right) $. A
characteristic feature is the positive curvature,
$d^{2}H_{p}\left( T\right) /dT^{2}>0$, for $\nu >1/2$. This
anomalous behavior has been found in a variety of cuprate
superconductors\cite{mackenzie,ando} and close to $T_{c}$ the data
appears to be consistent with $2\nu \approx 4/3$, the value for
3D-XY critical behavior (Eq.(\ref{eq19})).

To summarize, we have shown that in type II superconductors
subject to a magnetic field the correlation lengths cannot grow
beyond the magnetic length scale $L_{H}$. Invoking the finite size
scaling theory and the scaling form of the of the scaling form of
the free energy density of an anisotropic type II superconductor
close to zero field criticality we determined $L_{H}$ and explored
the resulting finite size effect. In contrast to its inhomogeneity
induced counterpart, arising from inhomogeneities with length
scale $L$, $L_{H}\propto H^{-1/2}$ can be varied continuously in
terms of the magnetic field strength. Thus, as long as $L>L_{H}$
the magnetic field induced finite size effect is observable and
allows to assess the importance of fluctuations, to extract
critical point properties of the homogeneous system and to derive
a lower bound for the length scale of inhomogeneities which affect
thermodynamic properties. Our analysis of specific heat data for
under- and optimally doped YBa$_{2}$Cu$_{3}$O$_{7-\delta }$, where
the critical regime is attained, revealed remarkable agreement
with the finite size scaling scenario. On the other hand the
existence of the limiting magnetic length scale does have even
away from the regime where critical fluctuations dominate
essential consequences. Since the correlation length $\xi ^{T}$
cannot grow beyond $L_{H}$, there is no critical line
$H_{c2}\left( T\right) $ retaining the continuous character of the
zero field transition. To illustrate this point we considered
MgB$_{2} $, 2H-NbSe$_{2}$ and Nb$_{77}$Zr$_{23}$, type II
superconductors with comparatively large correlation lengths.
Although the analyzed experimental data do not extend to the
critical regime, we have shown that there is a magnetic field
induced finite size effect and with that fluctuations at work.
Furthermore, lower bounds for the inhomogeneities affecting the
thermodynamic properties have been derived. Their length scales
range from $182$ A to $814$ A. Since the available data does not
extend to low fields, much larger values are conceivable. This
raises serious doubts on the relevance of the nanoscale spatial
variations in the electronic characteristics observed with
scanning tunnelling microscopy\cite {liu,chang,cren,lang}. In any
case, we have shown that the various predictions we have made here
are experimentally verifiable. In particular, the analogy with
finite size scaling is robust and it should be possible to
determine in exactly what regions the data collapses onto a
one-variable scaling formulation and to what extent corrections to
scaling can describe any deviations. Furthermore, in spite of the
consensus that in type II superconductors in an applied magnetic
field a phase transition to a state of zero resistance should
occur, we have shown that the magnetic finite size effect points
to the opposite conclusion. This is in agreement with the work
reported in Refs.\cite{strachan} and \cite{lobb}.

\bigskip

We thank H. Angst, H. Keller and C. Meingast for useful comments
and suggestions.
\end{document}